\documentclass[12pt, a4paper,oneside, reqno]{amsart}

\usepackage{amssymb,amsmath}
\usepackage{graphicx,psfrag, color}

\definecolor{hatter}{cmyk}{0,0,0.2,0}
\definecolor{kerdes}{cmyk}{0,0.2,0.1,0} 
\definecolor{valt}{cmyk}{0.1,0,0,0}   
\definecolor{def}{cmyk}{0.1,0,0.1,0}   
\definecolor{grey}{gray}{0.8}
\definecolor{ggrey}{gray}{0.9}
\definecolor{atirni}{gray}{0.8}
\newcommand{\comment}[1]{}

\newcommand{\Ob}{\mathrm{Ob}}
\newcommand{\Ph}{\mathrm{Ph}}
\newcommand{\vp}{\vec{p}}
\newcommand{\vq}{\vec{q}}
\newcommand{\vx}{\vec{x}}
\newcommand{\vy}{\vec{y}}
\newcommand{\va}{\vec{a}}

\newcommand{\vo}{\vec{o}}
\newcommand{\vz}{\vec{z}}
\newcommand{\vvkm}{\vec{v}^{\,k}_m}

\newcommand{\vakm}{\vec{a}^{\,k}_m}
\newcommand{\Q}{\mathrm{Q}}
\newcommand{\B}{\mathrm{B}}
\newcommand{\W}{\mathrm{W}}

\newcommand{\IOb}{\mathrm{IOb}}

\newcommand{\Time}{\mathsf{time}}
\newcommand{\dist}{\mathsf{dist}}
\newcommand{\leteq}{\mbox{$\,:=\,$}}
\newcommand{\upp}{\uparrow\hspace*{-1pt}\uparrow\!}
\newcommand{\scom}{\,\succ\hspace*{-1em}\prec}
\newcommand{\seq}{\equiv_{\,s}}
\newcommand{\simrad}{\thicksim^{rad}}
\newcommand{\simph}{\thicksim^{ph}}
\newcommand{\simmu}{\thicksim^\mu}
\newcommand{\then}{\enskip \Longrightarrow\ }
\newcommand{\rship}{\mbox{$>\hspace{-6pt}\big|$}b,k,c\mbox{$\big>_{\!rad}$}}
\newcommand{\mship}{\mbox{$>\hspace{-6pt}\big|$}b,k,c \mbox{$\big>_{\!\!\mu}$}}
\newcommand{\ship}{\mbox{$>\hspace{-6pt}\big|$}b,k,c \big>}
\renewcommand{\and}{\;\land\;}
\newcommand{\setclose}{\}}
\newcommand{\Setclose}{\,\right\}}
\newcommand{\setopen}{\{}
\newcommand{\Setopen}{\left\{\,}

\newcommand{\Dom}{Dom\,}

\usepackage{amsthm}
\usepackage{color}
\definecolor{thmcolor}{rgb}{0,0,.4}
\definecolor{remarkcolor}{rgb}{0,.2,0}
\definecolor{proofcolor}{rgb}{.4,0,0}
\definecolor{quecolor}{rgb}{.2,.2,0}
\definecolor{axcolor}{rgb}{.23,0,.23}

\newcommand{\ax}[1]{\textcolor{axcolor}{\ensuremath{\mathsf{#1}}}}

\theoremstyle{definition} \newtheorem{thm}{\textcolor{thmcolor}{Theorem}}[section]
\theoremstyle{definition} \newtheorem{col}[thm]{\textcolor{thmcolor}{Corollary}}
\theoremstyle{definition} 
\theoremstyle{definition} 

\theoremstyle{remark} \newtheorem{conv}[thm]{{\sc \textcolor{remarkcolor}{Convention}}}
\theoremstyle{remark} 
\theoremstyle{definition} 
\theoremstyle{definition}

\begin{document}

\title[FOL foundation of Relativity Theories]{First-order logic foundation of relativity theories}

\author{Judit X.\ Madar\'asz, Istv\'an N\'emeti and Gergely Sz\'ekely}

\date{2006. Apr. 10.}

\begin{abstract}
\comment{The aim of this paper is to give motivation and perspective
for working on a first-order logic foundation of relativity theories
(including general relativity). We present some of our axiom systems
and results in this area.} Motivation and perspective for an
exciting new research direction interconnecting logic, spacetime
theory, relativity---including such revolutionary areas as black
hole physics, relativistic computers, new cosmology---are presented
in this paper. We would like to invite the logician reader to take
part in this grand enterprise of the new century. Besides general
perspective and motivation, we present initial results in this
direction.
\end{abstract}

\maketitle

\section{Introduction (logic and spacetime geometry)}%
\label{geom-sec}

Throughout their intimately intertwined histories, logic and
geometry immensely profited from their interactions. In particular,
logic greatly profited from its applications to geometry. Indeed,
the very birth of logic was brought about by the needs of geometry
in the times of Socrates, Euclid and their predecessors. Ever since,
their interactions had rejuvenating, invigorating effects on logic.
For brevity, here we mention only Hilbert's axiomatization of
geometry, Tarski's improvements on this in the framework of
first-order logic (FOL)  \cite{Ta51}, Tarski's school of FOL
approaches to geometry as a small sample. It is no coincidence that
Tarskian algebraic logic is geometrical in spirit.

In this paper we try to show that this fruitful cooperation promises
new blessings for logic. This is so because there are breathtaking
revolutions in our understanding of space and time, i.e.\ in
relativity, cosmology, and black hole physics.

What is the subject matter of geometry? Traditionally, geometry was
created as a mathematical theory of a physical entity called space.
But recent developments in spacetime theory/general relativity show
that there is no such thing as physical space. Space is only an
illusion and as such is subjective. Space is a ``slice" of a larger
entity called spacetime. Spacetime, on the other hand, is objective,
it exists. What is subjective about space is the, necessarily ad
hoc, way we decide to ``slice" spacetime up into spacelike slices.
Actually, it was logician Kurt G\"odel who first discovered and
emphasized that in certain non-negligible cases such slicing is
impossible (non-foliazibility, in the technical terminology)
\cite{Godel}.

So, a great challenge for logic and logicians is to continue the
tradition sketched  above of providing foundation and conceptual
analysis for geometry by doing the same to spacetime theory, hence
to relativity.

A further motivation for geometry-friendly logicians is the
following. Relativity theory can be conceived of geometrizing parts
of physics in a sense, cf.\ \cite{MTW}. Special relativity (SR)
geometrizes some basic aspects of motion (kinematics) including
light propagation; general relativity (GR) geometrizes gravitation +
SR; the Kaluza-Klein style extension of GR geometrizes
electromagnetic phenomena + GR; and currently intensively researched
extensions of GR (e.g.\ string theory) search for extending the
scope of this aim for geometrizing more and more aspects of our
understanding of the world.

Why is this interesting for logicians? Well, because history tells
us that logic is applicable to geometry in an essential way. Hence
if relativity (and its extensions) is the act of geometrizing more
and more of physics, then it also can be regarded as a potential act
of ``logicizing" these areas, inviting logicians to take part in
this grandiose adventure of mankind.

\section{More concrete introduction (foundation of spacetime)}%

The idea of elaborating the foundational analysis of the logical
structure of spacetime theory and relativity theories (foundation of
relativity) in a spirit analogous with the rather successful
foundation of mathematics was initiated by several authors including
David Hilbert~\cite{Hi02},  cf.\ also Hilbert's 6th problem
\cite{Hi00}, Patrick Suppes~\cite{Sup59}, Alfred Tarski~\cite{HTS}
and leading contemporary logician Harvey Friedman~\cite{FriFOM1,
FriFOM2}.

There are several reasons for seeking an axiomatic foundation of a
physical theory \cite{Sup68}. One is that the theory may be better
understood by providing a basis of explicit postulates for the
theory. Another reason is that if we have an axiom system we can ask
ourselves what axioms are responsible for which theorems. For more
on this kind of foundational thinking called reverse mathematics,
see for example, Friedman~\cite{FriFOM1} and Simpson~\cite{Simpson}.
Furthermore, if we have an axiom system for special relativity or
general relativity, we can ask what happens with the theory if we
change one or more of the axioms. This could lead us to a new
physically interesting theory. This is what happened with Euclid's
axiom system for geometry when Bolyai and Lobachevsky altered the
axiom of parallelism and discovered hyperbolic geometry.

Seeking a logical foundation for spacetime theory (i.e., roughly,
relativity) is a worthwile attempt for several reasons. One of these
is that spacetime can be regarded as a foundation of physics since
spacetime is the arena in which physical phenomena take place.
Another reason for seeking a logical foundation for spacetime is
that throughout its history, logic benefited the most from those
applications of logic which were aiming at branches of learning
going through a turmoil or a revolutionary phase, and at the same
time being important for our understanding of the world
\cite{Hirsch}. As a quick glance to recent issues of, e.g.,
Scientific American can convince the reader, spacetime theory and
relativity/cosmology certainly qualify. So we believe that it serves
the best interest of logic community to apply logic to spacetime
theory, relativity, cosmology, and black hole physics. Indeed, logic
can benefit from such studies in many ways. As a bonus, as indicated
in \cite{Dgy-N} or \cite{HogarthPhD}, spacetime theory can give a
feedback to the foundation of mathematics itself.

For certain reasons, the foundation of mathematics has been carried
through strictly within the framework of first-order logic (FOL).
One of these reasons is that staying inside FOL helps us to avoid
tacit assumptions. Another reason is that FOL has a complete
inference system while higher-order logic cannot have one by
G\"odel's incompleteness theorem, see for example,
V{\"a}{\"a}n{\"a}nen~\cite[p.505]{V01}. For more motivation for
staying inside FOL as opposed to higher-order logic, see for
example,  \cite{AMNsamp}, \cite[Appendix 1: ``Why exactly
FOL'']{pezsgo}, \cite{Ax}, \cite{FeBSL}, \cite{Pam}, \cite{Wol}. The
same reasons motivate the effort of keeping the foundation of
spacetime and relativity theory inside FOL.

The interplay between logic and relativity theory goes back to
around 1920 and has been playing a non-negligible role in works of
researchers like Reichenbach, Carnap, Suppes, Ax, Szekeres,
Malament, Walker, and of many other contemporaries. For more
details, cf.~e.g., \cite{AMNsamp}. Also, it is no coincidence that
relativity was the main motivating example for the logical
positivists of the Vienna Circle.

Axiomatizations of SR have been quite extensively  studied in the
literature, see for example, the references of \cite{AMNsamp}.
However, these works usually stop with a kind of representation
theorem for their axiomatizations. As a contrast, what we call the
foundation of relativity begins with the axiomatization (and
representation theorems), but the real work and the real fun (the
conceptual analysis) comes afterwards when we investigate, e.g.,
what axioms are responsible for which statements, what happens if we
change the axioms etc.

While some FOL axiomatizations of the theory of inertial observers
and for SR can be found in the literature (\cite{Ax},
\cite{Goldblatt}, \cite{AMNsamp}), axiom systems---let alone FOL
axiom systems---for accelerated observers and for GR are not too
many in the literature (but cf.\ \cite{Twp} for an exception).

In section \ref{ax-sec}, we recall a streamlined FOL axiomatization
\ax{AccRel} of SR  extended with accelerated observers. In section
\ref{thm-sec}, we take one step toward GR and investigate an aspect
of time warp, that is the effect of gravitation on clocks, in our
FOL setting. There we use Einstein's equivalence principle to talk
about gravitation and prove the gravitational time dilation effect,
that is  that ``gravity causes time to run slow'', from \ax{AccRel}
in more than one sense. See Theorems \ref{thm-rad}, \ref{thm-mu} and
\ref{thm-ph}. We will also see that gravity can
 slow time down arbitrarily, see Theorems \ref{thm-Rad}, \ref{thm-Mu}
and \ref{thm-Ph}. Furthermore, we investigate the role of the
``direction'' and the ``magnitude'' of gravitation in gravitational
time dilation, see \ Theorems \ref{thm-ob} and \ref{thm-behind}. We
note that the most exotic features of black holes, wormholes and the
like (mentioned in section~\ref{why-sec} below) can be traced back
to this effect of time warp (to be analyzed in
section~\ref{thm-sec}).

\section{Intriguing features of GR spacetimes (challenges for the logician)}%
\label{why-sec}

Both SR and GR have many interesting consequences. Most of them show
that we have to refine our common sense concepts of space and time.
They are full of surprising predictions and paradoxes which
seriously challenge our common sense picture of the world. But it is
exactly this negation of common sense which makes this area an
attractive field to apply logic.

Gravitation has many surprising effects on time. The common name for
these effects is {\em time warp}.

For example, in the Schwarzschild spacetime, which is  associated
with a non-rotating black hole (or star), we face one of the
simplest aspect of time warp called {\em gravitational time
dilation}. There we see that if we suspend an observer closer to the
black hole and another observer farther away from it, then the clock
of the closer one will run slower than the clock of the one which is
farther away. So in some sense we see that ``gravity causes time run
slow''. There are places where this time warp effect becomes
infinite, i.e.\ some clocks entirely stop ticking, i.e.\ freeze from
the point of view of some other observers. Moreover, time and space
may get interchanged. These effects are part of the reason why we
said in section~\ref{geom-sec} that space does not exist while
spacetime does.

The above-mentioned time warp effect leads to even stronger effects.
We meet new interesting aspects of time warp in the
Reissner\,--\,Nordstr\"om, Kerr and Kerr\,--\,Newman spacetimes that
are associated with charged, rotating and charged-rotating black
holes, respectively. For astronomical evidence for the existence of
rotating black holes cf.~e.g., \cite{kerr1}, \cite{kerr2}. In these
spacetimes, there is an event whose causal past contains timelike
curves which are infinitely long in the future direction. Such a
curve can be the life-line of an observer (or computer) who has
infinite time for working and sending light-signals that can be
received before the distinguished event. The spacetimes in which
these kinds of events occur are called Malament\,--\,Hogarth
spacetimes, cf.~e.g., \ Earman~\cite[\S 4]{Earman},
\cite{HogarthPhD}. In Malament\,--\,Hogarth spacetimes, we can
design a computer that decides non-Turing computable sets, cf.~e.g.,
\cite{HogarthPhD}, \cite{Earman-Norton}, \cite{Dgy-N}, \cite{EN}.
Thus inside these spacetimes, we can decide whether an axiom system
of set theory (for example ZFC) is consistent or not. Therefore, in
contrast with the consequence of G\" odel's second incompleteness
theorem, we can find out whether mathematics is consistent or not.
For more detail on these kinds of computers in the physically
reasonable Kerr spacetime, cf.~e.g., \cite{Dgy-N}, \cite{EN}.
Recently, the acceleration of the expansion of the universe made
anti-de-Sitter spacetimes very popular with cosmologists. These also
have the Malament-Hogarth property, hence are also suitable for
harboring computers breaking the Turing barrier.

There are several models of GR in which there are so-called Closed
Timelike Curves (CTC). Such are G\" odel's rotating universe
\cite{Godel}, Kerr and Kerr\,--\,Newman spacetimes \cite{O'Neill},
Gott's spacetime \cite{Gott91}, Tipler's rotating cylinder
\cite{Tipler}, van Stockum's spacetime \cite{Stockum}, Taub-NUT
spacetime \cite{Hawking-Ellis}, to mention only a few.
Since timelike curves correspond to possible life-lines of
observers, in these spacetimes an observer can go through the same
event more than once. This situation can be interpreted as {\em time
travel}. This leads to non-trivial philosophical problems, in
analysing/understanding which the methods of logic can considerably
help. We believe, currently logic is the discipline best positioned
for clarifying the apparent problems with CTC's, i.e.\ with time
travel. Namely, the only problem with time travel is that it
represents a kind of circularity, because of the following: a time
traveler goes back into his past, changes his past so as to prevent
his own existence, but then who went back into the past? etc. This
circularity is not more vicious than the Liar paradox or
self-reference implemented e.g.\ in G\"odel's second incompleteness
proof. Logic has been extremely successful in understanding and
``de-mistifying'' self-referential situations and the Liar paradox.
Examples are provided by literature of G\"odel's incompleteness
method \cite{HajekPudlak}, the book on ``The Liar" by Barwise and
Etchemendy \cite{Liar} which used non-well-founded set theory for
providing an explicit semantic analysis for self-referential
situations, \cite{Sereny}. So logic seems to be best suited for
providing rational understanding of situations like the circularity
represented by CTC's or time travel. For more on CTC's, cf.~e.g.,
\cite[\S 6]{Earman}, \cite{ESW}, \cite{Gott}.

These are only a few of the many examples that show that turning
Relativity Theory into a real FOL theory, axiomatizing it and
analyzing its logical structure seem to be a promising, worthwile
undertaking.

What could science gain from such a logical analysis of relativity
theory? Turning GR into a FOL theory will make it more flexible. By
flexibility we mean that we can change some of the axioms whenever
we would like to change the theory, without having to re-build the
whole theory from scratch. By changing the axioms, we can control
the changes of theory better than by changing Einstein's field
equations. This might be useful when we would like to understand the
connection of GR to other theories of gravitation like the
Brans\,--\,Dicke theory, cf.~\cite{Brans97}, \cite{Brans05},
\cite{Faroni}. This flexibility can also be useful when we would
like to extend GR. We indeed would like to extend GR since we do not
have a good theory of Quantum Gravity (QG) which is a common
extension of the quantum theory and GR. Some eminent researchers of
relativity formulated an even more optimistic goal of searching for
the geometrization of all physical phenomena known today into a
so-called theory of everything (TOE). Of course, one wants both QG
and TOE to be some kinds of extensions of GR.

Recent astronomical observations provided strong evidence that the
expansion of our universe is accelerating. This discovery leads to
many questions and to the idea that the cosmological constant might
be replaced with a dynamical parameter, i.e.\ with a scalar field,
called Quintessence or ``dark energy"  cf.~e.g.,
\cite{Carroll},\cite{DGY}. But this leads to a new need for
modifying or at least fine-tuning GR. This also shows the merit of
making GR more flexible by providing a FOL axiom system for it.

So far we have talked mainly about the significance of the logical
foundation of GR, but the logical analysis of SR is also important
since GR is built on SR. Moreover, there are other different
relativity theories such as the Reichenbach\,--\,Gr\"unbaum version,
cf.~\cite{Reichenbach}, \cite{Reichenbach2} and \cite{Grunbaum} or
the  Lorentz-Poincar\'e version of special relativity
cf.~\cite{Lorentz}. Their logical structures and connection with
Einstein's relativity are also worth analyzing in order to get a
more refined understanding of relativity theory. Our research group
has done some work in this direction \cite[\S 4.5]{pezsgo}.

In the following sections we try to give a sample of the work done
by our research group in Budapest in the direction of a FOL
investigation of relativity theories (including GR).

\section{A FOL axiom system of SR extended with accelerated observers}%
\label{ax-sec}

Here we recall one of our axiom systems for SR extended with
accelerated observers (hence extended with a handle on gravity). We
try to be as self contained as possible. First occurrences of
concepts used in this work are set in boldface to make them easier
to find.

The motivation for our choice of vocabulary is summarized as
follows. Here we deal with the kinematics of relativity only, that
is we deal with motion of {\em bodies} (or {\em test-particles}). We
will represent motion as changing spatial location in time. To do
so, we will have reference-frames for coordinatizing events and, for
simplicity, we will associate reference-frames with special bodies
which we will call {\em observers}. We visualize an
observer-as-a-body as ``sitting'' in the origin of the space part of
its reference-frame, or equivalently, ``living'' on the time-axis of
the reference-frame. We will distinguish {\em inertial} observers
from non-inertial (accelerated) ones. There will be another special
kind of bodies which we will call {\em photons}. For coordinatizing
events, we will use an arbitrary {\em ordered field} in place of the
field of the real numbers. Thus the elements of this field will be
the ``{\em quantities}'' which we will use for marking time and
space. Allowing arbitrary ordered fields in place of the field of
the reals increases flexibility of our theory and minimizes the
amount of our mathematical presuppositions, cf.~e.g., Ax~\cite{Ax}
for further motivation in this direction. Similar remarks apply to
our flexibility oriented decisions below, for example, keeping the
dimension of spacetime a variable. Using observers in place of
coordinate systems or reference frames is only a matter of didactic
convenience and visualization. Using observers (or coordinate
systems, or reference-frames) instead of a single
observer-independent spacetime structure has many reasons. One of
them is that it helps us in weeding out unnecessary axioms from our
theories; but we state and emphasize the equivalence/duality between
observer-oriented and observer-independent approaches to relativity
theory, cf.\ \cite[\S 4.5]{Mphd}. Motivated by the above, we now
turn to fixing the first-order language of our axiom systems.

We fix a natural number $d\ge 2$ for the dimension of spacetime. Our
language contains the following non-logical symbols:
\begin{itemize}
\item unary relation symbols $\B$ (for {\bf Bodies}), $\Ob$ (for {\bf Observers}),
$\IOb$ (for {\bf Inertial Observers}), $\Ph$ (for {\bf Photons}) and $\Q$ (for {\bf Quantities}),
\item binary function symbols  $+$, $\cdot$ and a binary relation symbol $\le $ (for the field operations and the ordering on
$\Q$), and
\item a $2+d$-ary relation symbol $\W$ (for {\bf World-view relation}).
\end{itemize}

The bodies will play the role of the
``main characters''  of our spacetime models
and they will be ``observed'' (coordinatized using the
quantities) by the observers. This observation will be coded by the
world-view relation $\W$. Our bodies and observers are basically the same as
the ``test particles'' and the ``reference-frames'', respectively, in some of the literature.

We read $\B(x)$, $\Ob(x)$, $\IOb(x)$, $\Ph(x)$ and $\Q(x)$ as ``$x$ is a body'',
``$x$ is an observer'', ``$x$ is an inertial observer'', ``$x$ is a photon'', ``$x$ is a
quantity''.
We use the world-view  relation $\W$ to talk about coordinatization, by reading
$\W(x,y,z_1,\ldots, z_d)$ as ``observer $x$ observes (or sees)
body $y$ at coordinate point
$\langle z_1,\ldots,z_d\rangle$''.
This kind of observation has no connection with seeing via photons, it simply means
coordinatization.

$\B(x)$, $\Ob(x)$, $\IOb(x)$, $\Ph(x)$, $\Q(x)$, $\W(x,y,z_1,\ldots, z_d)$, $x=y$ and $x\leq y$
are the so-called {atomic formulas} of our first-order
language, where $x,y,z_1,\dots,z_d$ can be arbitrary variables or terms built
up
from variables by using the field-operations ``$+$'' and ``$\cdot$''. The
{\bf formulas} of  our first-order language are built up from these
atomic formulas by using the logical connectives {\em not}
($\lnot$), {\em and} ($\land$), {\em or} ($\lor$), {\em implies}
($\Longrightarrow$), {\em if-and-only-if} ($\Longleftrightarrow$) and the
quantifiers {\em exists} $x$ ($\exists x$) and {\em for all $x$} ($\forall x$)
for every variable $x$.

The {\bf models} of this language are  of the form
\begin{equation*}
\mathfrak{M} = \langle U; \B, \Ob, \IOb, \Ph, \Q,+,\cdot,\leq,\W\rangle,
\end{equation*}
where $U$ is a nonempty set and $\B$, $\Ob$, $\IOb$, $\Ph$ and $\Q$ are unary relations on $U$, etc.
A unary relation on $U$ is just a subset of $U$.
Thus we   use $\B,\Ob$ etc.\ as sets as well, for example, we  write $m\in \Ob$ in place of $\Ob(m)$.

$\Q^d\leteq\Q\times\ldots\times \Q$ ($d$-times) is the set of all $d$-tuples of elements of $\Q$.
If  $\vp\in \Q^d$, then we  assume that $\vp=\langle p_1,\ldots,p_d\rangle$, i.e.\ $p_i\in\Q$ denotes the $i$-th component of the $d$-tuple $\vp$.
We write $\W(m,b,\vp\,)$ in place of
$\W(m,b,p_1,\dots,p_d)$,  and we  write $\forall \vp$ in place of
$\forall p_1,\dots,p_d$ etc.

Let us begin formulating our axioms.  We formulate each axiom at two
levels. First we give an intuitive formulation, then we give a
precise formalization using our logical notation (which easily can
be translated into first-order formulas by substituting the
definitions into the formalizations). We aspire to formulate easily
understandable axioms in FOL.

The first axiom expresses our very basic assumptions like: both
photons and observers are bodies, inertial observers are also
observers, etc.

\begin{description}
\item[\ax{AxFrame}] $\Ob\cup \Ph\subseteq \B$, $\IOb\subseteq \Ob$,
$U=\B\cup \Q$, $\B\cap \Q=\emptyset$, $\W\subseteq \Ob \times
\B\times \Q^d$, $+$ and $\cdot$ are binary operations on $\Q$, $\le$
is a binary relation on $\Q$.
\end{description}

To be able to add, multiply and compare measurements of observers,
we put some algebraic structure on the set of quantities $\Q$ by the
next axiom.

\begin{description}
\item[\ax{AxEOF}]
A FOL axiom stating that the {\bf quantity part} $\left< \Q;
+,\cdot, \le \right>$ is a Euclidean\footnotemark\ ordered field.
\footnotetext{That is a linearly ordered field in which positive
elements have square roots.}
\end{description}
\noindent For the first-order definition of linearly ordered field,
see for example, Chang\,--\,Keisler~\cite{Chang-Keisler}.

\begin{figure}[h!btp]
\small
\begin{center}
\psfrag{mm}[bl][bl]{$tr_m(m)$}
\psfrag{mk}[br][br]{$tr_m(k)$}
\psfrag{mb}[t][t]{$tr_m(b)$}
\psfrag{mph}[tl][tl]{$tr_m(ph)$}
\psfrag{kk}[br][br]{$tr_k(k)$}
\psfrag{km}[b][bl]{$tr_k(m)$}
\psfrag{kb}[t][t]{$tr_k(b)$}
\psfrag{kph}[tl][tl]{$tr_k(ph)$}
\psfrag{p}[r][r]{$\vp$}
\psfrag{evm}[tl][tl]{$ev_m$}
\psfrag{evk}[bl][bl]{$ev_k$}
\psfrag{Evm}[r][r]{$Ev_m$}
\psfrag{Evk}[r][r]{$Ev_k$}
\psfrag{Ev}[r][r]{$Ev$}
\psfrag{T}[l][l]{$e=ev_m(\vp\,)=ev_k(\vq\,)$}
\psfrag{q}[r][r]{$\vq$}
\psfrag{Cdk}[l][l]{$Cd_k$}
\psfrag{Cdm}[l][l]{$Cd_m$}
\psfrag{Crdk}[l][l]{$Crd_k$}
\psfrag{Crdm}[l][l]{$Crd_m$}
\psfrag{t}[lb][lb]{$$}
\psfrag{o}[t][t]{$\vo$}
\psfrag{fkm}[t][t]{$f^k_m$}
\psfrag*{text1}[cb][cb]{world-view of $k$}
\psfrag*{text2}[cb][cb]{world-view of $m$}
\includegraphics[keepaspectratio, width=\textwidth]{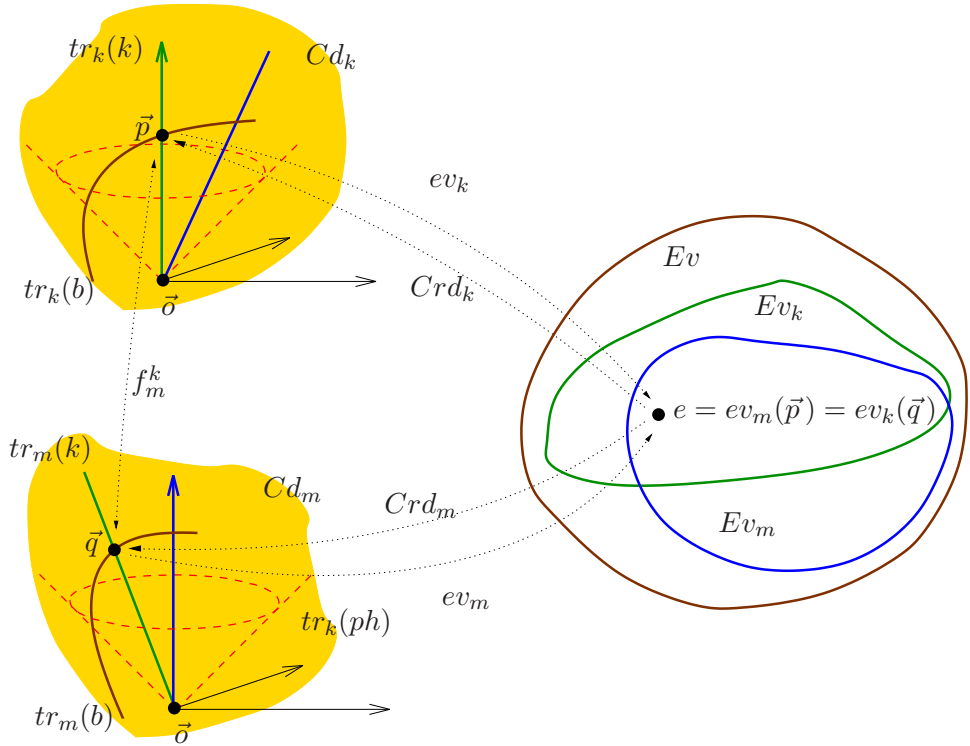}
\caption{\label{figfmk} Illustration for the basic definitions.}
\end{center}
\end{figure}

We will need some definitions to formulate our other axioms. Let
$0,1,-,/,\sqrt{\phantom{i}}$ be the usual field operations which are
definable from ``$+$'' and ``$\cdot$''. We use the vector-space
structure of $\Q^d$,  i.e.\ if $\vp,\vq\in \Q^d$ and $\lambda\in
\Q$, then $\vp+\vq,-\vp, \lambda \vp\in \Q^d$; and $\vo\leteq\langle
0,\ldots,0\rangle$ denotes the {\bf origin}. $\Q^d$ is called the
{\bf coordinate system} and its elements are referred to as {\bf
coordinate points}. We use the notation   $\vp_s\leteq\langle
p_2,\ldots, p_d\rangle$ for  the {\bf space component} of $\vp$ and
$p_t\leteq p_1$ for the {\bf time component} of $\vp\in\Q^d$. The
{\bf event}  (the set of bodies) observed by observer $m$  at
coordinate point $\vp$ is:
\begin{equation*}
ev_m(\vp\,)\leteq\Setopen b\in \B \::\: \W(m,b,\vp\,)\Setclose.
\end{equation*}
The {\bf coordinate-domain}  of  observer $m$ is the set of coordinate points where $m$
observes something:
\begin{equation*}
Cd_m\leteq\Setopen \vp \in \Q^d\::\: ev_m(\vp\,)\neq \emptyset  \Setclose.
\end{equation*}

Now we formulate our first axiom on observers. This natural axiom
goes back to Galileo Galilei and even to d'Oresme of around 1350,
cf.\ e.g., \cite[p.23, \S 5]{AMNsamp}. It simply states that each
observer thinks that he rests in the origin of the space part of his
coordinate system.

\begin{description}
\item[\ax{AxSelf^-}] An observer sees himself in an event iff the space component of the coordinate of this event is the origin.
\begin{equation*}
\forall m \in \Ob \enskip \forall \vp\in Cd_m\quad  \big(\,m\in ev_m(\vp\,) \iff \vp_s=\vo\;\big).
\end{equation*}
\end{description}

To formulate our axiom about the constancy of the speed of photons,
for convenience, we choose  $1$ for this speed. Below, the
{\bf Euclidean-length} of $\vp\in \Q^n$ is defined as $|\vp\,|
\leteq\sqrt{\resizebox{!}{8pt}{$p_1^2+\ldots+p_n^2$}}$, for any $n\ge 1$.
\begin{description}
\item[\ax{AxPh_0}]  For every \emph{inertial} observer, there is a photon through two
coordinate points $\vp$ and $\vq$ iff the slope of $\vp-\vq$ is $1$:
\begin{equation*}
\begin{split}
\forall m\in \IOb\enskip \forall \vp,\vq\in \Q^d\ \big(\,&
|\vp_s-\vq_s|=|p_t-q_t| \iff \\ &\Ph\cap ev_m(\vp\,)\cap
ev_m(\vq\,)\ne\emptyset \,\big).
\end{split}
\end{equation*}
\end{description}
Motivations for this axiom can be found, for example, in
\cite{logst}, or in d'Inverno~\cite[\S 2.6]{d'Inverno}.

The 
set of events seen by observer $m$ is:
\begin{equation*}
Ev_m\leteq\Setopen ev_m(\vp\,) \::\:  \vp\in Cd_m\Setclose,
\end{equation*}
and the set of all events is
\begin{equation*}
Ev\leteq\Setopen e\in Ev_m \::\:  m\in \Ob\Setclose.
\end{equation*}

With the next axiom, we assume that every inertial observer sees the same set of events.
\begin{description}
\item[\ax{AxEv}] Every \emph{inertial} observer sees the same events:
\begin{equation*}
\forall m,k\in \IOb \quad Ev_m=Ev_k.
\end{equation*}
\end{description}

One can prove from \ax{AxPh_0} and \ax{AxEOF} that if $m$ is an
\emph{inertial} observer and $e\in Ev_m$, then there is a unique coordinate
point $\vp\in\Q^d$ such that $e=ev_m(\vp\,)$. We will denote this
unique coordinate point $\vp\in\Q^d$ by $Crd_m(e)$.

\begin{conv}\label{Crdconv}
Whenever we write ``$Crd_m(e)$'', we mean that there is a unique
$\vq \in Cd_m$ such that $ev_m(\vq\,)=e$, and $Crd_m(e)$ denotes
this unique $\vq\,$. That is, if we talk about the value $Crd_m(e)$,
we postulate that it exists and is unique (by the present
convention).
\end{conv}
We say that events $e_1$ and $e_2$ are {\bf simultaneous}
\label{sim} for observer $m$, in symbols $e_1\sim_m e_2$, iff $e_1$
and $e_2$ have the same time-coordinate in $m$'s coordinate-domain,
i.e.\ if $Crd_m(e_1)_t=Crd_m(e_2)_t$.  To talk about time
differences measured by observers, we use $\Time_m(e_1,e_2)$ as an
abbreviation for $|Crd_m(e_1)_t-Crd_m(e_2)_t|$ and we call it the
{\bf elapsed time} between events $e_1$ and $e_2$ measured by
observer $m$. We note that, if $m\in e_1\cap e_2$, then
$\Time_m(e_1,e_2)$ is called the {\em proper time} measured by $m$
between  $e_1$ and $e_2$, and $e_1\sim_m e_2$ iff
$\Time_m(e_1,e_2)=0$. We use $\dist_m(e_1,e_2)$ as an abbreviation
for $|Crd_m(e_1)_s-Crd_m(e_2)_s|$ and we call it the {\bf spatial
distance} of events $e_1$ and $e_2$ according to an observer $m$. We
note that when we write $\dist_m(e_1,e_2)$ or $\Time_m(e_1,e_2)$, we
assume that $e_1$ and $e_2$ have unique coordinates by
Convention~\ref{Crdconv}.

\begin{description}
\item[\ax{AxSimDist}]If events $e_1$ and $e_2$ are simultaneous for both \emph{inertial} observers $m$ and $k$, then $m$ and $k$ agree on the spatial distance between $e_1$ and $e_2$:
\begin{equation*}
\begin{split}
\forall m,k\in \IOb\enskip \forall e_1,e_2 \in Ev_m \quad \big(&\,e_1\sim_m e_2 \and  e_1\sim_k e_2 \then \\
&\dist_m(e_1,e_2)=\dist_k(e_1,e_2)\,\big).
\end{split}
\end{equation*}
\end{description}

Let us collect these axioms in an axiom system called
\ax{SpecRel_\mathit{d}}.
\[
\ax{SpecRel_\mathit{d}}\leteq\setopen \ax{AxFrame}, \ax{AxEOF},
\ax{AxSelf^-}, \ax{AxPh_0}, \ax{AxEv},\ax{AxSimDist} \setclose.
\]
Now for each natural number $d\ge2$, we have a FOL theory of SR.
Usually we omit the dimension parameter $d$. From the few axioms
introduced so far, we can deduce the most frequently quoted
predictions, called paradigmatic effects, of SR: (i) ``moving clocks
slow down'' (ii) ``moving meter-rods shrink'' (iii) ``moving pairs
of clocks get out of synchronism''. For more detail, see for
example, \cite{AMNsamp}, \cite{pezsgo}, or \cite{logst}. Here we
concentrate on the behavior of clocks and indicate a connection with
Minkowski geometry.
\goodbreak

\begin{thm} \label{mink-thm}
Assume $\ax{SpecRel_\mathit{d}}$, $d\ge 3$. Then
$$\Time_m(e_1,e_2)^2-\dist_m(e_1,e_2)^2 =
\Time_k(e_1,e_2)^2-\dist_k(e_1,e_2)^2$$ \noindent for any
$m,k\in\IOb$ and $e_1,e_2\in Ev_m$.
\end{thm}

The above theorem is the starting point for building Minkowski
geometry, which is the ``geometrization" of SR. It also indicates
that time and space are intertwined in SR. Here we only concentrate
on its corollary usually stated as ``moving clocks slow down".
Theorem~\ref{mink-thm} shows that \ax{SpecRel} is a good axiom
system for SR if we restrict our interest to \emph{inertial} motion.

\begin{col}(moving clocks slow down) \label{srslow-cor}
Assume $\ax{SpecRel_\mathit{d}}$, $d\ge 3$. Let $m,k\in\IOb$,
$e_1,e_2\in Ev_k$,  and assume $k\in e_1\cap e_2$,
$\dist_m(e_1,e_2)\ne 0$. Then
$$\Time_m(e_1,e_2)>\Time_k(e_1,e_2).$$
\end{col}

In the above corollary,  a ``moving clock" is represented by
observer $k$, that he is moving relative to $m$ is expressed by
$\dist_m(e_1,e_2)\ne 0$, $k\in e_1\cap e_2$, and that $k$'s time is
slowing down relative to $m$'s is expressed by
$\Time_m(e_1,e_2)>\Time_k(e_1,e_2)$. This ``clock slowing down"
effect is only relative, i.e., ``clocks moving relative to $m$ slow
down relative to $m$". But this relative effect leads to a new kind
of gravitation-oriented ``absolute slowing time down" effect, as our
next theorem as well as the whole of section~\ref{thm-sec} will
show.

To extend \ax{SpecRel}, we now formulate axioms about non-inertial
observers. The non-inertial observers are called {\bf accelerated
observers}. Note that \ax{AxSelf^-} is the only axiom introduced so
far that talks about non-inertial observers, too. We assume the
following very natural axiom for all observers.

\begin{description}
\item[\ax{AxEv^+}]  Whenever an observer participates in an event, he also sees this event:
\begin{equation*}
\forall m\in \Ob\enskip \forall e\in Ev \quad \big(\,m\in e \then
e\in Ev_m\,\big).
\end{equation*}
\end{description}

\noindent The set of positive elements of $\Q$ is denoted by
$\Q^+\leteq\setopen x\in \Q:x>0\setclose$. The {\bf interval}
between $x,y\in \Q$ is defined as $(x,y)\leteq\{z\in\Q:x<z<y\}$. Let
$H\subseteq\Q$. We say that $H$ is {\bf connected} iff $\forall
x,y\in H\enskip (x,y)\subseteq H$, and we say that $H$ is {\bf open}
iff $\forall x\in H\enskip\exists \varepsilon\in \Q^+\enskip
(x-\varepsilon,x+\varepsilon)\subseteq H$.

We assume the following technical axiom:
\begin{description}
\item[\ax{AxSelf^+}]
The set of time-instances in which an observer is present in its own
world-view is connected and open:
\begin{equation*}
\forall m\in\Ob\enskip \{ p_t : m\in ev_m(\vp\,)\}\quad \mbox{is
connected and open.}
\end{equation*}
\end{description}

To connect the coordinate-domains of the accelerated and the
inertial observers, we are going to formulate the statement that at
each moment of his life, each accelerated observer sees the nearby
world for a short while as an inertial observer does. To formalize
this, first we introduce the relation of being a co-moving observer.
To do so, we define the  (coordinate) {\bf neighborhood} of event
$e$ with radius $r \in \Q^+$ according to observer $k$ as:
\begin{equation*}
B^r_k(e)\leteq\Setopen \vp\in Cd_k \::\: \exists \vq \in Cd_k \quad
ev_k(\vq\,)=e \and |\vp-\vq\,|<r\Setclose.
\end{equation*}
We note that $B^r_k(e)=\emptyset$ if $e\not \in Ev_k$ by this
definition. Observer $m$ is a {\bf co-moving observer} of observer
$k$ at event  $e$, in symbols $m \succ_e k$, iff the following
holds:
\begin{equation*}
\forall \varepsilon \in \Q^+ \;  \exists \delta \in \Q^+ \enskip
\forall \vp \in B^{\delta}_k(e)\quad
\big|\vp-Crd_m(ev_k(\vp\,))\big| \leq\varepsilon|\vp-Crd_k(e)|.
\end{equation*}
Note that $Crd_m(e)=Crd_k(e)$ and thus also $e\in Ev_m$ if $m
\succ_e k$ and $e\in Ev_k$. Note also that $m\succ_e k$ for every
observer $m$ if $e\not\in Ev_k$, by definition. Behind the
definition of the co-moving observers is the following intuitive
image: as we zoom into smaller and smaller neighborhoods of the
coordinate point of the given event, the coordinate-domains of  the
two observers are more and more similar. This intuitive picture is
symmetric while the co-moving relation $\succ_e$ is not. Thus we
introduce a symmetric version. We say that observers $m$  and $k$
are {\bf strong co-moving observers} at event $e$, in symbols $m
\scom_e k$, iff both $m\succ_e k$ and $k \succ_e m$ hold. The
following axiom gives the promised connection between the
coordinate-domains of the inertial and the accelerated observers:

\begin{description}
\item[\ax{AxAcc^+}] At any event in which an observer sees himself, there is a strong co-moving \emph{inertial} observer.
\begin{equation*}
\forall k \in \Ob \enskip \forall e \in Ev \quad (\,k\in e \then \exists m\in \IOb \enskip m \scom_e k\,).
\end{equation*}
\end{description}

The axioms introduced so far are not strong enough to prove
properties of accelerated clocks like the Twin Paradox, cf.\
Theorems 3.5 and 3.7 and Corollary 3.6 in \cite{Twp}. The additional
property we need is that every bounded non-empty subset of the
quantity part has a supremum. This is a second-order logic property
(because it concerns all subsets) which we cannot use in a FOL axiom
system. Instead, we will use a kind of ``induction'' axiom schema.
It will state that every non-empty, bounded subset of the quantity
part which can be defined by a FOL-formula using possibly the extra
part of the model, e.g., using the world-view relation, has a
supremum. To formulate this FOL induction axiom schema, we need some
more definitions.

If $\varphi$ is a formula and $x$ is a variable, then we say that
$x$ is a {\bf free variable} \label{free variable} of $\varphi$ iff
$x$ does not occur under the scope of either $\exists x$ or $\forall
x$. Sometimes we introduce a formula $\varphi$ as $\varphi(\vx\,)$,
this means that all the free variables of $\varphi$ lie in $\vx$.

If $\varphi(x,y)$ is a formula and $\mathfrak{M}=\langle
U;\ldots\rangle$ is a model, then whether $\varphi$ is true or false
in $\mathfrak{M}$ depends on how we associate elements of $U$ to the
free variables $x,y$. When we associate $a,b\in U$ to $x,y$,
respectively, then $\varphi(a,b)$ denotes this truth-value, thus
$\varphi(a,b)$ is either true or false in $\mathfrak{M}$. For
example, if $\varphi$ is $x\leq y$, then $\varphi(0,1)$ is true
while $\varphi(1,0)$ is false in any ordered field. A formula
$\varphi$ is said to be {\bf true} in $\mathfrak{M}$ if $\varphi$ is
true in $\mathfrak{M}$ no matter how we associate elements to the
free variables. We say that a {\bf subset $H$ of $\Q$ is}
(parametrically) {\bf definable by} $\varphi(y,\vx\,)$ iff there is
$\va\in U^n$ such that $H=\setopen b\in \Q\: :\:
\varphi(b,\va\,)\text{ is true in }\mathfrak{M}\setclose$. We say
that a subset of $\Q$ is {\bf definable} iff it is definable by a
FOL-formula.

Let $\phi(x,\vy\,)$ be a FOL-formula of our language.
\begin{description}
\item[\ax{AxSup_\phi}] Every subset of $\Q$ definable by  $\phi(x,\vy\,)$ has  a supremum if it is non-empty and {\bf bounded}.
\end{description}
\noindent A FOL formula expressing \ax{AxSup_\phi} can be found in
\cite{Twp}. Our axiom scheme \ax{IND} below says that every
non-empty bounded subset of $\Q$ that is definable in our language
has a supremum:
\begin{equation*}
\ax{IND}\leteq\Setopen \ax{AxSup_\varphi}\::\:    \varphi \text{ is a FOL-formula of our language} \Setclose.
\end{equation*}
Note that \ax{IND} is true in any model whose quantity part is the
field of real numbers. For more detail about \ax{IND}
cf.~\cite{Twp}.

Let us call the collection of the axioms introduced so far \ax{AccRel_\mathit{d}}:
\begin{equation*}
\begin{split}
\ax{AccRel_\mathit{d}}\leteq\ax{SpecRel_\mathit{d}}\cup\Setopen\ax{AxEv}^+,
\ax{AxSelf}^+, \ax{AxAcc^+}\Setclose\cup\ax{IND}.
\end{split}
\end{equation*}

The so-called Twin Paradox is provable in \ax{AccRel},
cf.~\cite{Twp}, \cite{mythes}. We now formulate the Twin Paradox
with our logical notation.

The {\bf set of events encountered by $m\in \Ob$ between $e_1,e_2\in
Ev$} is denoted as
\begin{equation*}
\begin{split}
Ev_m(e_1,e_2)\leteq &\Setopen e\in Ev_m\::\: m\in e \and\right. \\
&\left. Crd_m(e_1)_t<Crd_m(e)_t<Crd_m(e_2)_t \Setclose.
\end{split}
\end{equation*}

Now we can formulate the Twin Paradox in our FOL setting.

\begin{description}
\item[\ax{TwP}] Every \emph{inertial} observer $m$ measures more time
than or equal time as any other observer $k$ between any two meeting
events $e_1$ and $e_2$; and they measure the same time iff they have
encountered the same events between $e_1$ and $e_2$:
\begin{equation*}
\begin{split}
\forall e_1,e_2\in Ev\enskip\forall m \in \IOb\enskip \forall  k\in \Ob &\quad\Big(\, k,m\in e_1\cap e_2 \then\\
\big(\,\Time_m(e_1,e_2)=\Time_k(e_1,e_2)&\iff Ev_m(e_1,e_2)=Ev_k(e_1,e_2) \,\big) \\
& \and\Time_m(e_1,e_2)\ge\Time_k(e_1,e_2)\Big).
\end{split}
\end{equation*}
\end{description}

The following theorem states that the Twin Paradox is provable in
\ax{AccRel_\mathit{d}} if $d\ge3$.

\begin{thm}
 \ax{AccRel_\mathit{d}}\,\,$\models$ \ax{TwP},\quad if $d\ge 3$.
\end{thm}

\noindent
For the proof of this theorem, cf.~\cite{Twp}, \cite{mythes}.
\medskip

We note that there are non-trivial models of \ax{AccRel}. E.g., the
construction in Misner\,--\,Thorne\,--\,Wheeler~\cite[\S 6,
especially pp.\ 172-173 and \S 13.6 on pp.\ 327-332]{MTW} can be
used for constructing models for \ax{AccRel}.

\section{One step toward GR (effect of gravitation on clocks)}%
\label{thm-sec}

We would like to investigate the effect of gravitation on clocks in
our FOL setting. As a first step we prove theorems about the
Gravitational Time Dilation that roughly says that ``gravitation
makes time flow slower'', that is, the clocks in the bottom of a
tower run slower than the clocks in the top of the tower. We will
use Einstein's equivalence principle to treat gravitation in
\ax{AccRel}. This principle says that a uniformly accelerated frame
of reference is indistinguishable from a rest frame in a uniform
gravitational field, cf.~e.g., d'Inverno~\cite[\S 9.4]{d'Inverno}.
So instead of gravitation we will talk about acceleration and
instead of towers we will talk about spaceships. This way the
Gravitational Time Dilation will become the following statement:
``the time in the aft of an accelerated spaceship flows slower than
in the front of the spaceship''. We begin to formulate this
statement in our FOL language.

\begin{figure}[h!]
\small
\begin{center}
\psfrag{a}[tl][tl]{$tr_m(k)$}
\psfrag{m}[l][l]{$tr_m(m)$}
\psfrag{e}[b][b]{$e$}
\psfrag{e1}[tl][tl]{$e_1$}
\psfrag{e'}[bl][bl]{$e'$}
\psfrag{e2}[br][br]{$e_2$}
\psfrag{2l}[l][l]{$2\lambda$}
\psfrag{l}[b][b]{$\lambda$}
\psfrag{ph1}[tr][tr]{$ph_1$}
\psfrag{ph2}[br][br]{$ph_2$}
\psfrag{text1}[tl][tl]{$(a)$}
\psfrag{text2}[tl][tl]{$(b)$}

\includegraphics[keepaspectratio, width=\textwidth]{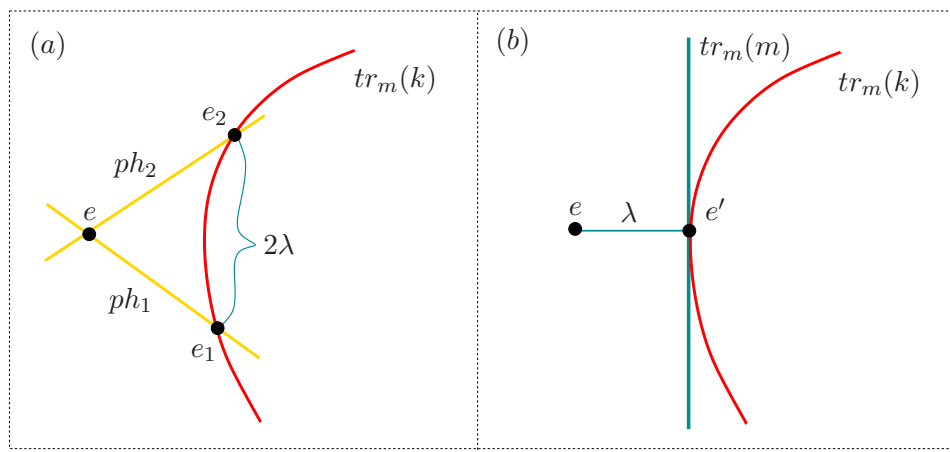}
\caption{\label{distfig} $(a)$ for the radar-distance and $(b)$ for
the Minkowski-distance.}
\end{center}
\end{figure}

To talk about spaceships, we will need a concept of distance between
events and observers. We have the following two natural candidates
for this:
\begin{itemize}
\item Event $e$ is at {\bf radar-distance} $\lambda\in\Q^+$ from observer $k$ iff there are events $e_1$ and $e_2$ and photons $ph_1$ and $ph_2$ such that $k\in e_1\cap e_2$, $ph_1\in e\cap e_1$, $ph_2\in e\cap e_2$ and $\Time_k(e_1,e_2)=2\lambda$.
Event $e$ is at {\bf radar-distance} $0$ from observer $k$ iff $k\in e$.
See $(a)$ of Figure \ref{distfig}.
\item Event $e$ is at {\bf Minkowski-distance} $\lambda\in\Q$ from observer $k$ iff there is an event $e'$ such that $k\in e'$, $e\sim_m e'$ and $\dist_m(e,e')=\lambda$  for every inertial co-moving observer $m$ of $k$ at $e'$.
See $(b)$ of Figure \ref{distfig}.
\end{itemize}

We say that  observer $k$ thinks that body $b$ is at constant radar
(Minkowski) distance from him iff the radar-distance
(Minkowski-distance) of every event which $b$ participates in is the
same.

The {\bf life-line}%
\footnote{Life-line is called world-line in some of the literature.}
(or {\bf trace}) of body $b$ according to observer $m$ is defined as
the set of coordinate points where $b$ was observed by $m$:
\begin{equation*}
tr_m(b)\leteq\Setopen \vp\in \Q^d \::\: \W(m,b,\vp\,)\Setclose.
\end{equation*}
\noindent Note that $tr_m(b)=\Setopen \vp\in \Q^d \::\: b \in
ev_m(\vp\,)\Setclose$. For stating that the {\em spaceship does not
change its direction} we introduce the following concept. We say
that observers $k$ and  $b$ are {\bf coplanar}  iff $tr_m(k)\cup
tr_m(b)$ is a subset of a plane containing a line parallel with the
time-axis, in the coordinate system of an {\em inertial} observer
$m$.

We now introduce two concepts for spaceships. Observers $b,k$ and
$c$ form a {\bf radar-spaceship}, in symbols $\rship$, iff $b$, $k$
and $c$ are coplanar and $k$ thinks that $b$ and $c$ are at constant
radar-distances from him. The definition of the {\bf
Minkowski-spaceship}, in symbols  $\mship$, is analogous.

We say that event $e_1$  (causally) {\bf precedes} event $e_2$
according to observer $k$ iff $Crd_m(e_1)_t \le Crd_m(e_2)_t$ for
all \emph{inertial} co-moving observers $m$ of $k$. In this case, we
also say that $e_2$ {\bf succeeds} $e_1$ according to $k$.

\begin{figure}[h!btp]
\small
\begin{center}
\psfrag{e}[b][b]{$e$}
\psfrag{e'}[b][b]{$e'$}
\psfrag{e1}[b][b]{$e_1$}
\psfrag{e2}[b][b]{$e_2$}
\psfrag{ph1}[b][b]{$ph_1$}
\psfrag{ph2}[b][b]{$ph_2$}
\psfrag{m}[lb][lb]{$m$}
\psfrag{a}[tl][tl]{$k$}
\psfrag{l}[b][b]{$\lambda$}
\psfrag*{text1}[cb][cb]{(a)}
\psfrag*{text2}[cb][cb]{(b)}
\psfrag*{text3}[cb][cb]{(c)}
\includegraphics[keepaspectratio, width=\textwidth]{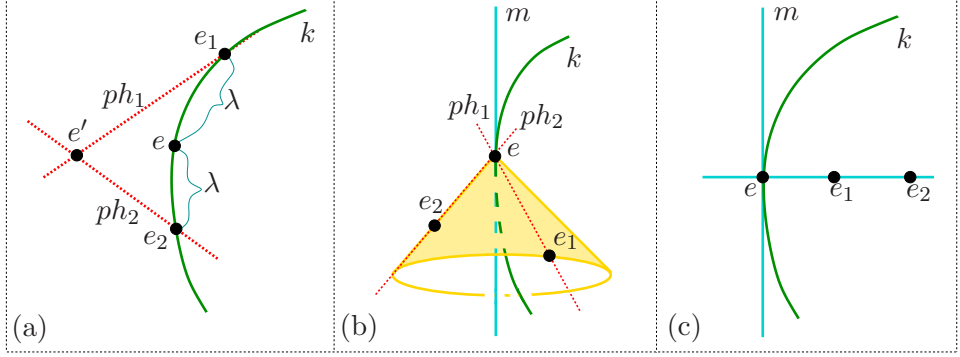}
\caption{\label{simfig} $(a)$ is for $e\simrad_k e'$, $(b)$ is for
$e_1\simph_k e_2$  and $(c)$ is for $e_1\simmu_k e_2$.}
\end{center}
\end{figure}

We need some concept for deciding which events happened at the same
time according to an accelerated observer. The following three
natural concepts offer themselves:
\begin{itemize}
\item Events  $e$ and $e'$  are {\bf radar-simultaneous} for observer $k$, in symbols $e\simrad_k e'$, iff  $k\in e$ and there are events $e_1$ and $e_2$ and photons $ph_1$ and $ph_2$ such that  $k\in e_1\cap e_2$, $ph_1\in e\cap e_1$, $ph_2\in e\cap e_2$ and $\Time_k(e_1,e)=\Time_k(e,e_2)$ or there is an event $e_3$ such that $e\simrad_k e_3$ and $e_3\simrad_k e'$.
See $(a)$ of Figure \ref{simfig}.
\item Events  $e_1$ and $e_2$  are {\bf photon-simultaneous} for observer $k$, in symbols $e_1\simph_k e_2$,  iff there is an event $e$ and photons $ph_1$ and $ph_2$ such that $k\in e$, $ph_1\in e\cap e_1$,  $ph_2\in e\cap e_2$ and  $e_1$ and $e_2$ precedes $e$ according to $k$.
See $(b)$ of Figure \ref{simfig}.
\item Events  $e_1$ and $e_2$  are {\bf Minkowski-simultaneous} for observer $k$, in symbols $e_1\simmu_k e_2$,  iff there is an event $e$ such that $k\in e$ and $e_1$ and $e_2$ are simultaneous for any inertial co-moving observer of $k$ at $e$.
See $(c)$ of Figure \ref{simfig}.
\end{itemize}
We note that, for \emph{inertial} observers, the concepts of
radar-simultaneity and Minkowski-simultaneity coincide with the
concept of simultaneity introduced on page \pageref{sim}.

\begin{figure}[h!btp]
\small
\begin{center}
\psfrag{p}[lb][lb]{$\vp$} \psfrag{ps}[lb][lb]{$\vp_s$}
\psfrag{qs}[lb][lb]{$\vq_s$} \psfrag{ph}[rb][rb]{$ph$}
\psfrag{q}[r][r]{$\vq$} \psfrag{q3}[l][l]{$\vq_3$}
\psfrag{q2}[l][l]{$\vq_2$} \psfrag{q1}[l][l]{$\vq_1$}
\psfrag{aa}[l][l]{$(a)$} \psfrag{bb}[l][l]{$(b)$}
\psfrag{o}[t][t]{$\vo$} \psfrag{b}[t][t]{$k$} \psfrag{k}[t][t]{$b$}
\psfrag{e}[t][t]{$e$} \psfrag{eb}[t][t]{$e_k$}
\psfrag{ek}[t][t]{$e_b$} \psfrag{k1}[t][t]{$b'$}
\psfrag{b1}[t][t]{$k'$}

\includegraphics[keepaspectratio, width=\textwidth]{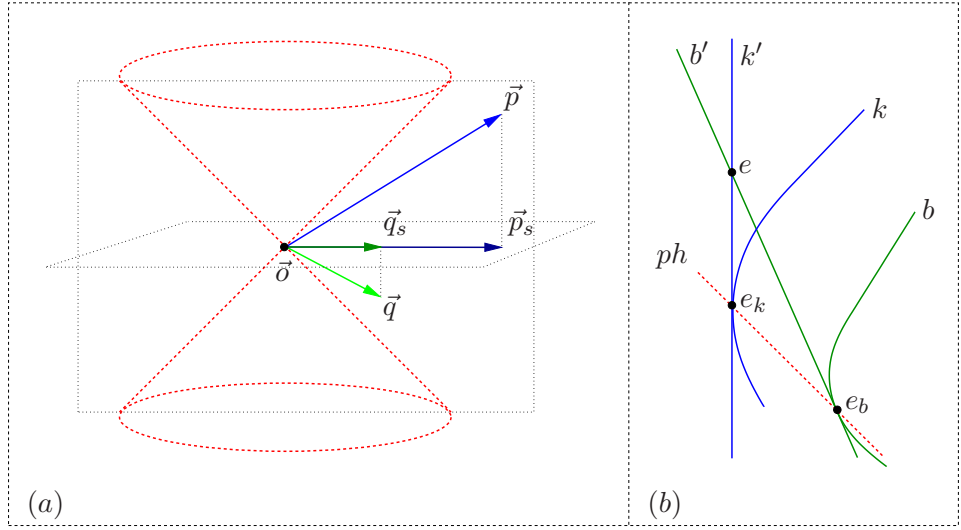}
\caption{\label{figupp} $(a)$ is for illustrating $\vp\upp\vq$ and
$(b)$ is for illustrating observer $b$ approaching to observer $k$,
as seen by $k$ with photons.}
\end{center}
\end{figure}

We will distinguish the front and the aft of the spaceship by the
direction of the acceleration. Thus we need a concept for direction.
We say that the {\bf directions of $\vp\in \Q^d$ and $\vq\in \Q^d$
are the same}, in symbols $\vp\upp \vq$, iff there is a $\lambda \in
\Q^+$ such that $\lambda \vp_s=\vq_s$, see $(a)$ of
Figure~\ref{figupp}.

Now let us turn our attention towards the definition of acceleration
in our FOL setting.

We define the {\bf life-curve} of observer $k$ according to observer
$m$ as the life-line of $k$ according to $m$ {\em parameterized by
the time measured by $k$}, formally:
\begin{equation*}
\begin{split}
Tr^k_m &\leteq\Setopen \langle t,\vp\, \rangle\in \Q\times Cd_m \::\right.\\
&\left.\exists \vq\in tr_k(k)\quad q_t=t\land
ev_m(\vp\,)=ev_k(\vq\,) \Setclose.
\end{split}
\end{equation*}
The {\bf domain} of a binary relation $R$ is defined as $\Dom
R\leteq\setopen x : \exists y \enskip \langle x,y\rangle \in
R\setclose$.

Both the life-curves of observers (according to any {\em inertial}
observer) and the derivative $f'$ of a given function $f$ are
first-order logic definable concepts, cf.~\cite{Twp}. Thus the
following definitions are also FOL ones: The {\bf relative-velocity}
$\vvkm$ of observer $k$ according to observer $m$ at instant
$t\in\Q$ is the derivative of the life-curve of $k$ according to $m$
at $t$, i.e., $\vvkm(t)=(Tr^k_m)'(t)$ if  $t\in \Dom Tr^k_m$ and
undefined otherwise. The {\bf relative-acceleration} $\vakm$  of
observer $k$ according to observer $m$ at instant $t\in\Q$ is the
derivative of the relative-velocity at $t$ if it is differentiable
at $t$ and undefined otherwise.

Events $e_1$ and $e_2$ are called {\bf spacelike separated}, in
symbols $e_1\seq e_2$, iff $Crd_m(e_1)$ and $Crd_m(e_2)$ can be
connected by a line of slope less than 1 for every {\em inertial}
observer $m$, i.e., iff
$|(Crd_m(e_1)-Crd_m(e_2))_s|>|(Crd_m(e_1)-Crd_m(e_2))_t|$ for every
inertial observer $m$. We say that {\bf the direction of the
spaceship $\ship$ agrees with that of the acceleration of $k$} iff
the following holds:
\begin{equation*}
\begin{split}
\forall & m \in \IOb \enskip \forall t \in \Dom \vakm\enskip \forall e_1,e_2 \in Ev \quad  \big(\,c\in e_1 \and\\
&b\in e_2\and e_1\seq e_2 \then  \vakm(t)\upp
(Crd_k(e_1)-Crd_k(e_2))\,\big).
\end{split}
\end{equation*}

The (signed) {\bf Minkowski-length} of $\vp\in \Q^d$ is
\begin{equation*}
\mu(p)\leteq\left\{
\begin{array}{rl}
\sqrt{\big|p_t^2-|\vp_s|^2 \big|}  & \text{ if }  p_t^2-|\vp_s|^2\ge0 , \\
-\sqrt{\big|p_t^2-|\vp_s|^2 \big|} & \text{ otherwise }
\end{array}
\right.
\end{equation*}
and the {\bf Minkowski-distance} between  $\vp$ and $\vq$ is
$\mu(\vp,\vq\,)\leteq\mu(\vp-\vq\,)$. A motivation for the
``otherwise'' part of the definition of  $\mu(\vp\,)$  is the
following. $\mu(\vp\,)$  codes two kinds of information, (i) the
length of  $\vp$  and (ii) whether  $\vp$  is timelike  (i.e.\
$|p_t|>|\vp_s|$) or spacelike. Since the length is always
non-negative, we can use the sign of $\mu(\vp\,)$  to code  (ii).

The {\bf acceleration} of an observer $k$ at instant $t\in\Q$ is
defined as the Minkowski-length of the relative-acceleration seen by
any {\em inertial} observer $m$ at $t$, that is:
\begin{equation*}
a_k(t)\leteq\mu\big(\vakm(t)\big).
\end{equation*}
The acceleration is a well defined concept since it is independent
of the choice of the inertial observer $m$. We say that  observer
$k$ is {\bf positively accelerated} iff $a_k(t)\neq 0$ for all $t\in
\Dom Tr^k_k$. Observer $k$ is called {\bf uniformly accelerated} iff
there is an $a\in\Q^+$ such that $a_k(t)=a$ for all $t\in \Dom
Tr^k_k$.

We say that {\bf the  clock of $b$ runs slower than the clock of $c$
as seen by $k$ with radar  (photons; Minkowski-simultaneity)} iff
$\Time_b(e_b,e'_b)<\Time_c(e_c,e'_c)$ for all events $e_b,e'_b, e_c,
e'_c$ for which $b\in e_b\cap e'_b$, $c\in e_c\cap e'_c$ and
$e_b\simrad_k e_c$, $e'_b\simrad_k e'_c$. ($e_b\simph_k e_c$,
$e'_b\simph_k e'_c$; $e_b\simmu_k e_c$, $e'_b\simmu_k e'_c$).

Now we can state our first theorem about the clock-slowing effect of
gravitation:

\begin{thm} \label{thm-rad}
Assume \ax{AccRel_\mathit{d}} and $d\ge 3$. Let $\rship$ be a radar-spaceship such that:
\begin{itemize}
\item[(1)] $k$ is positively accelerated,
\item[(2)] the direction of the spaceship agrees with that of the acceleration of
$k$.
\end{itemize}
$(i)$ Then the clock of $b$ runs slower than the clock of $c$ as seen by $k$ with radar and
$(ii)$  the clock of $b$ runs slower than the clock of $c$ as seen by each of $k$, $b$ and $c$ with photons.
\end{thm}

To state a similar theorem in Minkowski-spaceships, we need the
following concept. We say that observer {\bf $b$ is not too far
behind} positively accelerated observer $k$ iff the following holds:
\begin{equation*}
\begin{split}
\forall m \in \IOb \enskip &\forall t \in \Dom Tr^k_m\enskip \forall \vp,\vq \in Cd_m \quad \big(\,\vp\in tr_m(k)\and \\
\vq\in tr_m(b)&\and ev_m(\vp\,)\simmu_k ev_m(\vq\,) \and \vakm(t)\upp (\vp-\vq\,) \then \\
&\qquad\qquad\quad\forall \tau \in \Dom\vakm \quad \mu(\vp-\vq\,)<1/a_k(\tau)\,\big).
\end{split}
\end{equation*}

Now we can state our second theorem about the clock-slowing effect
of gravitation:
\begin{thm}\label{thm-mu}
Assume \ax{AccRel_\mathit{d}} and $d\ge 3$. Let $\mship$ be a Minkowski-spaceship such that:
\begin{itemize}
\item[(1)] $k$ is positively accelerated,
\item[(2)] the direction of the spaceship agrees with that of the acceleration of $k$,
\item[(3)] $b$ is not too far behind $k$.
\end{itemize}
Then $(i)$ the clock of $b$ runs slower than the clock of $c$ as
seen by $k$ with Minkowski-simultaneity or with photons and $(ii)$
the clock of $b$ runs slower than the clock of $c$ as seen by each
of $k$, $b$ and $c$ with photons.
\end{thm}

In the following theorem we will see that the flow of time as seen
by photons is strongly connected with the following two concepts. We
say that observer $b$ is {\bf approaching} to ({\bf moving away}
from) observer $k$ as seen by $k$ with photons iff for all events
$e_k$ and $e_b$, if $b\in e_b$, $k\in e_k$ and $e_k\simph_k e_b$,
then there is an event $e$ such that  $k',b'\in e$ for all inertial
co-moving observers $k'$ and $b'$ of $k$ and $b$ at events $e_k$ and
$e_b$, respectively, and $e_b$ precedes (succeeds) $e_k$ according
to $k$, cf.\  (b) of Figure \ref{figupp}.

We say that the life-curve of observer $k$ is continuously
differentiable if the curve $Tr^k_m$ is such for all inertial
observers $m$.

\begin{thm} \label{thm-ph}
Assume \ax{AccRel_\mathit{d}} and $d\ge 3$.
Let $b,k\in\Ob$ such that $b$ and $k$ are coplanar and the life-curve of $k$ is continuously differentiable.
\begin{itemize}
\item[(1)] If $b$ is approaching to $k$ as seen by $k$ with photons, then the clock of $k$ runs slower
than the clock of $b$ as seen by $k$ with photons.
\item[(2)] If $b$ is moving away from $k$ as seen by $k$ with photons, then the clock of $b$ runs
slower than the clock of $k$ as seen by $k$ with photons.
\end{itemize}
\end{thm}

None of the axioms introduced so far require the existence of
accelerated (non-inertial) observers. Our following axiom scheme
says  that every definable timelike curve is the life-line of an
observer. Since from \ax{AxSelf^-}, \ax{AxPh_0} and \ax{AxEv} it
follows that the life-lines of inertial observers are straight
lines, cf.~e.g., \cite{AMNsamp}, \cite{Ramon}, \cite{RSz}, this will
ensure the existence of many non-inertial observers.

A differentiable function $\gamma$ is called {\bf timelike curve}
iff the slope of $\gamma'(t)$ is less than $1$ (i.e.,
$|\gamma'(t))_s|<|\gamma'(t))_t|$) for all $t\in \Dom \gamma$ and
$\Dom \gamma$ is an open and connected subset of $\Q$. It is clear
that this is a first-order logic definable concept since every
fragment of it is such. We say that a {\bf function $f$ is}
(parametrically) {\bf definable by} $\psi(x,\vy, \vz\,)$ iff there
is  $\va\in U^n$ such that $f(b)=\vp \iff \psi(b,\vp,\vz\,)$ true in
$\mathfrak{M}$.

Let $\psi$ be a FOL-formula of our language.
\begin{description}
\item[\ax{Ax\exists Ob_\psi}] If a function parametrically definable by $\psi$ is a timelike curve,
then there is an observer whose life-line is  the range of this function.
\end{description}
\begin{equation*}\ax{Ax\exists Ob}\leteq\Setopen  \ax{Ax\exists Ob_\psi}\::\: \psi \text{
is a FOL-formula of our language } \Setclose.
\end{equation*}

The following three theorems say that the clocks can run arbitrarily
slow or fast, as seen with the three different methods.

\begin{thm} \label{thm-Rad}
Assume \ax{AccRel_\mathit{d}}, \ax{Ax\exists Ob} and $d\ge 3$.  Let
$k\in\Ob$ be positively accelerated such that $\Dom Tr^k_k=\Q$  and
let $e,e'\in Ev$ such that $e\neq e'$ and $m\in e\cap e'$. Then for
all $\lambda \in \Q^+$, there are an observer $b$ and events
$e_b,e'_b\in Ev$ such that $b\in e_b\cap e'_b$, $e\simrad_m e_b$,
$e'\simrad_m e'_b$ and $\Time_b(e_b,e_b')=\lambda \Time_m(e,e')$.
\end{thm}

\begin{thm}\label{thm-Mu}
Assume \ax{AccRel_\mathit{d}} and $d\ge 3$.  Let $k\in\Ob$ be
uniformly accelerated and let $e,e'\in Ev$ such that $e\neq e'$ and
$m\in e\cap e'$. Then for all $\lambda \in \Q^+$, there are an
observer $b$ and events $e_b,e'_b\in Ev$ such that $b\in e_b\cap
e'_b$, $e\simmu_m e_b$, $e'\simmu_m e'_b$ and
$\Time_b(e_b,e_b')=\lambda \Time_m(e,e')$.
\end{thm}

\begin{thm} \label{thm-Ph}
Assume \ax{AccRel_\mathit{d}}, \ax{Ax\exists Ob} and $d\ge 3$.  Let
$k\in\Ob$ be  positively accelerated and $e,e'\in Ev$ such that
$e\neq e'$ and $m\in e\cap e'$. Then for all $\lambda \in \Q^+$,
there are an observer $b$ and events $e_b,e'_b\in Ev$ such that
$b\in e_b\cap e'_b$, $e\simph_m e_b$, $e'\simph_m e'_b$ and
$\Time_b(e_b,e_b')=\lambda \Time_m(e,e')$.
\end{thm}

We have seen that gravitation (acceleration) makes ``time flow
slowly''. However, we left open the question what role the
``strength'' and the ``direction'' of the gravitation play in this
effect. The following theorem shows that two observers, say $m$ and
$k$, can feel the same gravitation while the clock of $k$ runs
slower than the clock of $m$. Thus it is not  the ``strength'' of
the gravitation that makes ``time flow more slowly''.

\begin{thm} \label{thm-ob}
Assume \ax{AccRel_\mathit{d}}, \ax{Ax\exists Ob} and $d\ge 3$. There
are uniformly accelerated observers $m$ and $k$ such that
$a_k(t)=a_m(t)$ for all $t\in\Q$, but the clock of $k$ runs slower
than the clock of $m$ as seen by both $m$ and $k$ with photons (or
with radar or with Minkowski simultaneity).
\end{thm}

Now let us see what we can say about the role of the ``direction''
of gravitation. Being ``more down in a gravitational well" becomes
being ``behind" if we translate it from the language of gravitation
into the language of acceleration. This can be formulated by our
notation as follows. We say that observer $b$ is {\bf behind}
observer $k$ iff
\begin{equation*}
\begin{split}
\forall m \in \IOb&\enskip \forall t \in \Dom Tr^k_m\enskip \forall \vp,\vq \in Cd_m\quad\vp\in tr_m(k) \and \\
&\vq\in tr_m(b)\and ev_m(\vp\,)\simmu_k ev_m(\vq\,) \and
\vakm(t)\upp (\vp-\vq\,).
\end{split}
\end{equation*}

The following theorem shows that if observer $b$ is at a lower level
in the tower than observer $k$ is, then his clock runs slower than
the clock of $k$, as seen by $k$ with radar.

\begin{thm} \label{thm-behind}
Assume \ax{AccRel_\mathit{d}}, \ax{Ax\exists Ob} and $d\ge 3$. Let
$b,k\in\Ob$ such that:
\begin{itemize}
\item[(1)] $k$ is positively accelerated,
\item[(2)] $b$ and $k$ are coplanar,
\item[(3)] $b$ is behind $k$.
\end{itemize}
Then the clock of $k$ runs slower than the clock of $b$, as seen by
$k$ with radar.
\end{thm}

The proofs, along with more explanation and motivation, of the
theorems presented in this section can be found in \cite{GTD}.

\section{Questions, suggestions for future research}
\label{quest-sec}

(1) We hope that the perspective outlined in
sections~\ref{geom-sec}-\ref{why-sec}, and the techniques presented
in sections~\ref{ax-sec}-\ref{thm-sec}, \cite{Twp} already suggest a
research proposal. Sections~\ref{ax-sec}-\ref{thm-sec} cover only a
small fragment of the research proposed in
sections~\ref{geom-sec}-\ref{why-sec}. So the proposal is: elaborate
a larger part of the perspective outlined in
\ref{geom-sec}-\ref{why-sec} in the style of
sections~\ref{ax-sec}-\ref{thm-sec} and \cite{Twp}.

(2) The Introduction of \cite{pezsgo} contains more ideas both on
the general perspective (of applying logic to spacetime theory) and
also more of the long-distance goals. However, some of the present
results were not available when \cite{pezsgo} was written, therefore
that introduction does not replace completely the present section.

(3) In section~\ref{thm-sec} we started to elaborate a purely
logical theory of the effects of gravitation on clocks. Elaborate
this direction in more detail, and investigate more aspects of
gravitation on clocks. E.g.\ assume we bore a hole through the Earth
from the North pole to the South pole. Now put a clock into the
middle of the Earth. It will levitate ``weightlessly" in the middle.
Put another clock to the surface of the Earth. It will be squeezed
by gravity to the surface. Despite  this, the clock levitating in
the middle will run slower than the one on the surface. A third
clock high above in deep space will run even faster (than the one on
the surface). Why? Find a logic style formulation of the above (and
prove it) in the manner of section~\ref{thm-sec}.

(4) Investigate/formulate further aspects of the effects of gravity
on instruments (like clocks, meter-rods). E.g.\ define the so-called
gravitational force-field experienced by an accelerated observer
(via acceleration, relative to the observer, of test particles
dropped by the observer). Study this force-field and connect this
study with the investigations in section~\ref{thm-sec}. Try to make
an integrated coherent picture of gravity, time warp (clock behavior
in gravitational fields), and gravitational force. (Remark:
gravitational force is often suppressed in the literature because it
is not ``absolute", i.e.\ is not observer independent. All the same,
if we keep in mind that it is observer dependent, then it is a
helpful concept.) Imagine a long, accelerated spaceship. The
gravitational force experienced in the aft of the ship will be
greater than that in the front of the ship. Why?

(5) Continuing in the spirit of sections~\ref{ax-sec}-\ref{thm-sec},
\cite{Twp}, and the above, elaborate a FOL theory of the spacetime
of a Schwarzschild black hole \cite{TW00}. Streamline that theory,
make it logically transparent and illuminating. Apply conceptual
analysis to the theory similar in spirit as conceptual analysis of
special relativity is started in
\cite{pezsgo},\cite{AMNsamp},\cite{MNT}. Using the theory of
accelerated observers and Einstein's equivalence principle, create a
logically convincing, illuminating theory of such black holes. In
this direction it might be helpful that the analogy between the
world-view or reference frame of an accelerated spaceship and
skyscrapers (towers) on the event horizon of a black hole is
described in detail in Rindler's relativity book \cite[\S 12.4 ``The
uniformly accelerated lattice", pp.267-272]{Rindler}. Figure~12.6 is
especially useful therein. Also note how in Rindler's arrangement of
the skyscrapers above the black hole they are prevented from falling
by rigid rods separating them (these rods provide the
``acceleration" experienced by the inhabitants of the
towers/spaceships). These rods are called struts in \cite[p.270,
line 7 bottom up]{Rindler}.

So, we suggest combining the presently started FOL theory of
accelerated observers and of effects of gravity (acceleration) on
instruments of observers with the just quoted part of Rindler's work
in order to elaborate a FOL theory of the simplest kind of black
holes. Of course, the main point is that we are striving for a very
special kind of illuminating (etc.) FOL theory (and not just any FOL
theory describing a black hole).

When the above is done, we suggest applying re-coordinatization in
order to obtain an Eddington-Finkelstein version of this FOL theory
of the black hole. This second (EF) version of the theory will also
describe what the in-falling observer sees e.g.\ from inside the
event horizon. For the latter question we suggest assuming that the
black hole is huge (galactic size) so that enough stuff remains to
be observed after falling through the event horizon.

(6) After having streamlined, analyzed, simplified FOL theories of
simple (but huge) black holes, we propose turning to what we call
double black holes or exotic black holes. Double black holes have
two event horizons, an outer one and an inner one. In theory and
under certain assumptions, a traveler might fall into the black
hole, survive this and may come out at some other point of spacetime
(in our universe or in some other universe). So, some of these
double black holes may be regarded kind of wormholes. Examples are
spinning black holes (Kerr spacetime, Kerr-Newmann spacetime), and
electrically charged black holes (Reissner-Nordstr\"om spacetime)
\cite{TW00}.

The task here is again to build up, streamline, and conceptually
analyse, simplify FOL theories for such double black holes. They
offer logically intriguing issues for the logician as indicated in
section~\ref{why-sec}.

(7) Besides the relatively simple kind of acceleration studied in
sections~\ref{ax-sec}-\ref{thm-sec}, \cite{Twp}, rotation provides a
kind of acceleration appearing in the form of the centrifugal force.
A further research task is to analyse via FOL the world-view
represented by a rotating coordinate lattice (relative to the
gyroscopes) and generally, the rotational spacetimes. An example for
these is the slowly rotating black hole (Kerr spacedtime), other
examples are G\"odel's rotating universe, Tipler - van Stockum
spacetime. In these spacetimes rotation leads to CTC's and to many
other exotic effects like the so-called dragging of inertial frames
or the drag effect. Finding out more about these is the task of
NASA's recent ``Probe B". Here again a FOL theory of such spacetimes
waits for the creation, conceptual analysis and detailed
illuminating explanation of what happens and exactly why. A
particular question waiting to be answered is to find out and
analyse what the common features/mechanisms/principles of these
rotating spacetimes (with CTC's) are. E.g., many features of the
above mentioned three spacetimes coincide. Is this a coincidence or
is there a more general ``theory of rotating spacetimes" lurking in
the background. For more on this question we refer to \cite{ANW}. In
particular, we are looking for a logical answer to the
quasi-philosophical question: ``Exactly why and how CTC's are
generated in rotating black holes and in G\"odel's universe. Why do
they counter-rotate with matter?" (More on what we call
``counter-rotation" can be found in \cite{ANW}.)

\section*{ACKNOWLEDGEMENTS}

Thanks go to Victor Pambuccian for many valuable conversations on the subject, for
reading an earlier version of this paper and for helpful suggestions leading
to the present version.

Research supported by the
        Hungarian National Foundation for scientific research grant
        T43242 as well as by Bolyai Grant for Judit X.\ Madar\'asz.

\bibliographystyle{plain}

\end{document}